\documentclass[twocolumn,showpacs,preprintnumbers,amsmath,amssymb]{revtex4}

\usepackage{dcolumn}
\usepackage{bm}
\usepackage{latexsym}
\usepackage{amsfonts}
\usepackage[dvips]{graphicx}
\usepackage[usenames]{color}
\usepackage{makeidx}

\newcommand{\ket}{\rangle }
\newcommand{\bra}{\langle }

\newcommand{\ve}{\varepsilon}

\newcommand{\up}{\uparrow}
\newcommand{\dw}{\downarrow}

\begin{document}


\title{Interfaces of correlated electron systems:  Proposed
mechanism for colossal electroresistance 
}

\author{Takashi Oka$^{1}$}
\author{Naoto Nagaosa$^{1,2,3}$}
\affiliation{
$^1$Correlated Electron Research Center (CERC),
National Institute of Advanced Industrial Science and Technology (AIST),
Tsukuba Central 4, Tsukuba 305-8562, Japan.\\
$^2$Department of Applied Physics, University of Tokyo,
Bunkyo-ku, Tokyo 113-8656, Japan.\\
$^3$CREST, Japan Science and Technology Corporation (JST),
Saitama, 332-0012, Japan.
}

\date{\today}
\begin{abstract}
\noindent 
Mott's metal-insulator transition at an
interface due to band bending is studied
by the density matrix renormalization group (DMRG). 
We show that the result can be recovered by 
a simple modification of the conventional 
Poisson's equation approach used in semi-conductor heterojunctions.
A novel mechanism of colossal electroresistance is proposed,
which incorporates the hysteretic behavior of the transition
in higher dimensions.
\end{abstract}

\maketitle

\maketitle
Strongly correlated electron system (SCES) -- 
a group of materials in which 
the effect of Coulomb repulsion 
is large -- is one of the major candidates 
on which the next-generation electronics may be built 
\cite{Asamitsu1997,Ponnambalam1999,Oshima1999,Liu2000,
Baikalov2003,Sawa2004,Hasegawa2004,OhtomoNature}.
This strong hope put on to SCES electronics stems from the richness 
of its phase diagram \cite{Imada1998}; 
Various electronic and magnetic phases 
transitions are reported with high sensitivity to external conditions. 
Inside electric devices, the electrons behave collectively,
so the high sensitivity of SCES may lead to drastic functionalities.

Yet, SCES electronics faces a strong conceptual 
barrier to widespread acceptance and application,
since it is essentially many-body
where the useful concepts such 
as band structure are believed to fail.
It is expected to be the case especially 
at the interfaces.
In this Letter, we study {\it interface Mott transition}
near an interface between 
a metal electrode and SCES
in terms of the density matrix renormalization group (DMRG) 
method \cite{dmrg}. 
Surprisingly, we found that the conventional 
band bending picture based on Poisson's equation 
is valid with a small modification 
(eq. (\ref{eq:Poisson})).
Namely, the conventional concepts in semiconductor devices are still
useful and valid to design the SCES devices.  
As an application, we propose a novel mechanism of the 
colossal electroresistance (CER), i.e.,
the large switching of resistance.

Among many possibilities to industrial applications, 
perhaps CER
in SCES heterostructures
is closest to realization, e.g., 
nonvolatile resistance random
access memories (RRAM) \cite{Asamitsu1997}.
The device consists of a film of perovskite manganite 
such as Pr$_{1-\delta}$Ca$_{\delta}$MnO$_3$, $\delta=0.3$ (PCMO), 
which is a holed-doped Mott insulator,  
sandwiched by two metallic electrodes.
The current-voltage curve shows 
large hysteresis at room temperature, where the 
resistivity of the on and off 
states differ by a large factor.
Although this effect has been explored in details
\cite{Ponnambalam1999,Oshima1999,Liu2000,Baikalov2003,Sawa2004}, 
the understanding of the mechanism is 
still missing.
Baikalov {\it et al.} pointed out from multileads 
resistance measurements, that the 
switching take place at the interface \cite{Baikalov2003}. 
Then, Sawa {\it et al.} reported that the 
CER behavior depends on the 
work-function of the electrode metal \cite{Sawa2004}.
They interpreted the $I-V$ characteristics
using a Schottky contact model (metal/p-type semiconductor) 
accompanied with an interface state. 
Another mechanism was proposed by 
Rozenberg {\it et al.} \cite{Rozenberg2004}. 

The mechanism of CER we propose here do not assume 
any interface states 
but attributes the 
large non-linearity of the 
$I$-$V$ characteristics 
to interface Mott transition, 
where a layer of Mott insulator blocks the current. 
In dimensions higher than one, the transition 
is first order, and the width of the Mott insulator layer 
depends on how the voltage is changed.
This explains the hysteretic behavior of the $I$-$V$ characteristics. 

\begin{figure}[t]
\centering 
\includegraphics[width=6cm]{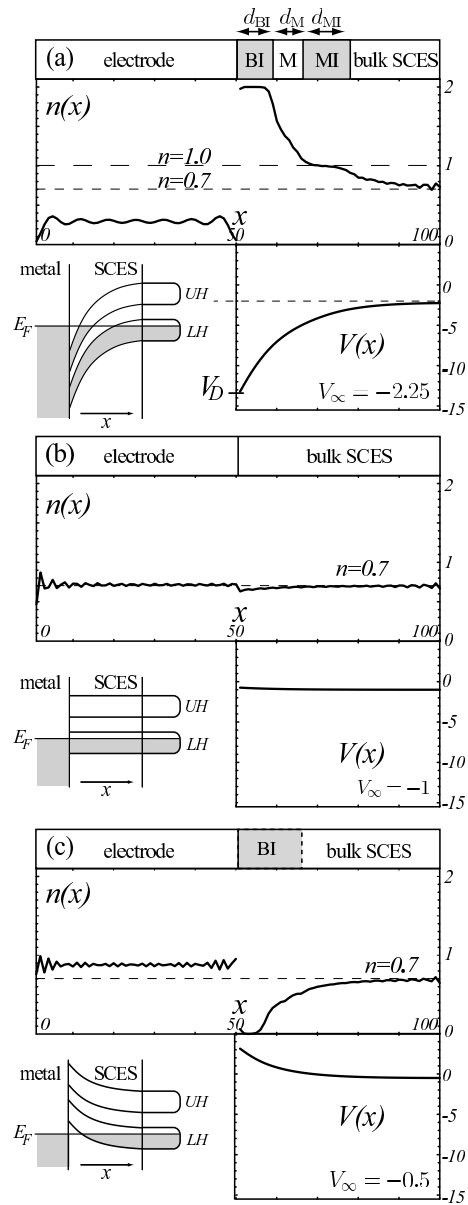}
\caption{
Electron density $n(x)$ (upper panel) and potential $V(x)$
(lower panel) in a metal/SCES interface with (a)
$V_{\infty}=-2.25$, (b) $V_{\infty}=-1$,
and (c) $V_{\infty}=-0.5$
 ($U/t=4$, $n_+=0.7$, $\alpha=0.043$ and $L=100$). 
In each panels, a description of 
each regions is given in the top where 
dark (white) regions are insulator (metal),
e.g. band insulator (BI),
metal (M), and Mott insulator (MI). 
The schematic picture in the left-lower space
describes the bending of the upper (lower) Hubbard band UH (LH)
near the interface. $E_F$ is the Fermi 
energy.
}
\label{fig:U4result}
\end{figure}

The Letter consists of two parts. 
We first study a one-dimensional 
model of a metal/SCES interface using DMRG
combined with Poisson's equation. 
We show that interface Mott transition can be understood 
on the basis of local equilibrium, i.e., 
electric state is determined by the local value of the potential. 
Then, we explore the hysteresis loop of the $I$-$V$ characteristics 
on a phenomenological basis
assuming the a hysteretic density-potential curve
for a system in higher dimensions.
Along with the mechanism of CER, another interesting 
consequence of interface Mott transition is proposed: 
A {\it quantum well} structure emerges 
spontaneously (Fig.\ref{fig:U4result}(a))
at the interface of a hole (electron) doped 
Mott insulator and an electrode with small (large) work-function. 
Our results may lead to fabrication of clean 2D metallic 
systems analogous to the high electron mobility 
transistor in semiconductor physics.

{\it DMRG study of a 1D interface ---}
We start with the one-dimensional model of 
a metal/SCES interface on a lattice 
with total Hamiltonian 
\begin{equation}
H_{\rm tot}=H_{\rm elc}+H_{\rm SCES}+H_{\rm jnc},
\end{equation}
where the electrode 
$H_{\rm elc}=-t\sum_\sigma\sum_{i=1}^{L/2-1}\left(c^\dagger_{i+1\sigma}
c_{i\sigma}+H.c\right)$
and SCES 
$H_{\rm SCES}=
-t\sum_\sigma\sum_{i=L/2+1}^{L-1}\left(c^\dagger_{i+1\sigma}
c_{i\sigma}+H.c\right)
\quad +\sum_{i=L/2+1}^{L}\left(Un_{i\up}n_{i\dw}+V_in_i\right)$
is connected by a junction
$H_{\rm jnc}=-t\sum_\sigma\left(c_{L/2+1\sigma}^
\dagger c_{L/2\sigma}+H.c.\right)$.
$t$ is the hopping element, $L$ the total size of the 
system with the interface at the center, 
and $U$ denotes the on-site Coulomb repulsion (for a related work, see 
\cite{Yonemitsu2005}).
We take the units $\hbar=\ve_0=1$. 
The potential $V_i$ defined in the SCES region obeys 1D Poisson's equation, 
whose discretised solution is 
\begin{eqnarray}
V_i=-\alpha\sum_{j=i}^{L}\sum_{l=j}^{L}
(n_l-n_+)+V_{\infty},\quad  i\in [L/2+1, L],
\label{eq:potential}
\end{eqnarray}
where
$n_l=\bra \sum_\sigma c^\dagger_{l\sigma}c_{l\sigma}\ket$
is the electron density and
$n_+$ the positive background
related to the hole doping ratio $\delta$ by $n_+=1-\delta$,
and $\alpha=\frac{e }{\ve a },\;(a:\mbox{lattice const.},\;
\ve: \mbox{dielectric constant})$.
$[a,b]$ stands for the interval between $a$ and $b$.
$H_{\rm SCES}$, without $V_i$,
is identical to the Hamiltonian of the one-dimensional Hubbard model, 
which exhibits a Mott 
transition at half-filling if $U>0$ \cite{Lieb:1968AM}. 
If $n\ne 1$, the ground-state is metallic, a state
known as the Tomonaga-Luttinger liquid
(e.g. \cite{Schulz1992}).

The DMRG calculation is performed as follows. 
In order to obtain the ground-state
consistent with the potential determined by Poissons's equation, 
we update the potential using eq.(\ref{eq:potential})
at each step of the finite size method in the 
DMRG procedure \cite{dmrg}.
This is repeated more than 
$50$ times to obtain total convergence.
The typical size of the block Hilbert space used here is $m=250$.
We fix the total number of 
electrons $\sum_{i=1}^Ln_i=Ln_+$, 
thus, the Fermi level of the electrode changes 
as electrons (holes) are injected to the SCES region
when we vary $V_\infty$. 
The interface is characterized by the
work-function difference 
$V_{\rm D}\equiv V_{L/2+1}-V_{L}=(\mbox{work-function of electrode})-
(\mbox{work-function of SCES})$.
We note that $V_\infty$ achiving a given $V_{\rm D}$
depends on $L$.

\begin{figure}[t]
\centering 
\includegraphics[width=8cm]{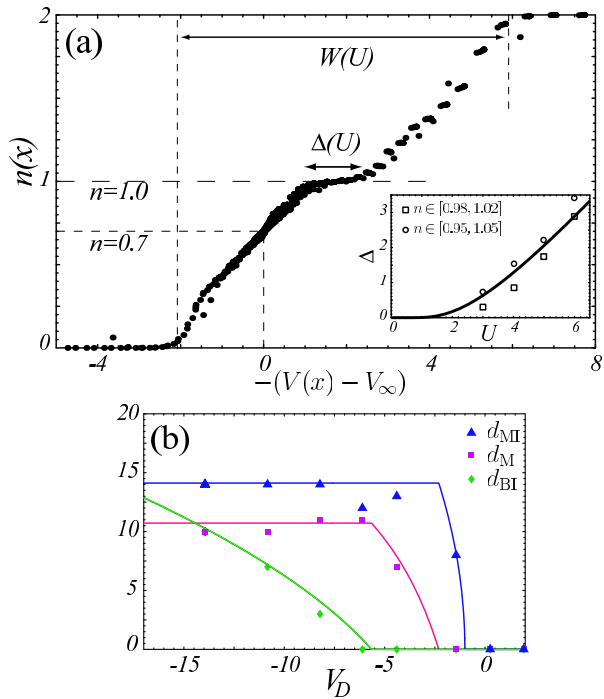}
\caption{
(a)
Universal density-potential relation
of the 1D interface Mott transition.
Electron density $n(x),\;51\le x\le 90$ is plotted against $-(V(x)-V_{\infty})$.
$V_{\infty}$ is varied from $V_{\infty}=-3.00$ to $V_{\infty}=0$ with an
interval of $0.25$. 
$W(U)$ is the band width and $\Delta(U)$ the 
width of the $n=1$ plateau.
(inset) The $U$ dependence of $\Delta(U)$. 
This is determined from data with density 
$n\in [0.96,1.04]$ (open circle) and $n\in [0.98,1.02]$ (open box).
The solid line is the Mott gap from Lieb-Wu's solution \cite{Lieb:1968AM}.
(b) The width of the Mott insulating 
($d_{\rm MI}$, $n\in [0.85,1.04]$; blue), metallic ($d_{\rm M}$,
$n\in [1.04,1.96]$; red) 
and band insulating ($d_{\rm BI}$, $n\in [1.96,2.0]$; green) 
layers plotted against $V_{\rm D}$.
The symbols are DMRG results 
while the solid
lines are the solutions of Poisson's equation (eqs. 
(\ref{eq:MIwidth})-(\ref{eq:BIwidth})).
}
\label{fig:NvsV}
\end{figure}

In Fig. \ref{fig:U4result}, we plot the electron 
density and the potential in a metal/SCES interface. 
Three typical solutions are plotted in a 
increasing order 
of $V_{\rm D}$ from (a) to (c).
\begin{description}
	\item[(a) Interface Mott transition:] 
	When the electrode's work-function is small enough, a
	quantum well is formed at the interface, i.e.,
	two insulating (MI and BI) layers with width $d_{\rm MI},\;
	d_{\rm BI}$, 
    sandwiches a metallic region (M) with width $d_{\rm M}$. 
    The $V_{\rm D}$ dependences of these widths are
    plotted in Fig. \ref{fig:NvsV} (b). 
	\item[(b) Ohmic junction:]
	The Fermi surface of the electrode and SCES balances 
	and no barrier is formed.
	\item[(c) Schottky barrier:]
	A Schottky barrier is formed as in conventional metal/semiconductor interfaces.
\end{description}
It is noted here that the qualitative features of the
results are all captured well by the conventional band bending
picture by replacing the valence (conduction) 
band by the lower (upper) Hubbard band as shown in Fig.\ref{fig:U4result}.
We also note that in the three cases, $n(x)$ 
shows an oscillatory behavior in the 
electrode regime, which is the 
1D Friedel oscillation
$\delta n(x)=\cos(2k_Fr+\eta_F)/r$
with $k_F$ the 
Fermi wave number, $\eta_F$ a phase shift,
and $r$ the distance from the interface.

In Fig.\ref{fig:NvsV} (a), we plot the electron density $n(x)$ against 
$-(V(x)-V_{\infty})$ in the SCES regime ($x\in [51,90]$),
where data near the boundary ($x\in [91,100]$) 
were omitted to avoid the boundary effect.  
The data falls onto a universal 
density-potential curve, which increases as the potential 
becomes deeper.
In the middle of the curve, there is a plateau at half-filling
whose width is $\Delta(U)$. 
The interface Mott transition is an 
analog of the filling driven Mott transition \cite{Fisher1995,Kotliar2002}.
There the filling $n(\mu)$ of a grand canonical system is studied 
while the chemical potential $\mu$ is varied. 
The universal density-potential curve 
is an analogue of the $n(\mu)$ curve in 
the filling driven Mott transition, 
where $-V(x)$, with a shift of the zero point, plays the role of $\mu$.
Indeed, when we compare the width of the 
plateau with the Lieb-Wu solution of the Mott gap 
of the one-dimensional Hubbard model
$
\Delta(U)=\frac{16t}{U}\int_1^{\infty}\frac{\sqrt{y^2-1}}{\sinh (2\pi yt/U)}dy
$ \cite{Lieb:1968AM},
the two coincides well as shown in the inset of Fig.\ref{fig:NvsV} (a).
Thus, it can be said that the density-potential 
relation in the interface Mott transition follows the 
$n(\mu)$ curve in the bulk transition.

{\it Poisson's equation and local equilibrium approximation ---}
Since the metal/SCES interface 
determines the transport properties of the device, 
the width of the layers
$d_{\rm MI}$, $d_{\rm M}$, and
$d_{\rm BI}$ in Fig. \ref{fig:U4result} (a)
is of practical interest.
Here, we derive
the width by solving the modified
Poisson's equation
\begin{equation}
\frac{d^2V(x)}{dx^2}=-\frac{e}{\ve}(n(V(x))-n_+),
\label{eq:Poisson}
\end{equation}
where we assume local equilibrium, 
that is, we assume that the electron density only
depends on the local value of the potential. 
In the following, we evaluate 
eq. (\ref{eq:Poisson}) to obtain 
the widths expressed solely by the 
potential difference $V_{\rm D}$, hole doping ratio $\delta$,
band width $W$, and the Mott gap $\Delta$.

We adopt a simplified density-potential relation
by linearizing the DMRG result Fig. \ref{fig:NvsV} (a); 
A constant compressibility $-dn(V)/dV=\kappa\equiv 2/(W-\Delta)$ 
is assumed for $-(V(x)-V_{\infty})\in [(\delta-1)/\kappa,\delta/\kappa],\;
[\delta/\kappa+\Delta,(1+\delta)/\kappa+\Delta]$,
and $n(V)=0$ for $-(V(x)-V_{\infty})\in[-\infty,(\delta-1)/\kappa]$,
$n(V)=1$ for 
$-(V(x)-V_{\infty})\in[\delta/\kappa,\delta/\kappa+\Delta]$ 
and
$n(V)=2$ for $-(V(x)-V_{\infty})\in[(1+\delta)/\kappa+\Delta,\infty]$.
We seek for a solution 
with a fixed density at 
the bulk SCES $n(x)=n_+=1-\delta$,
Mott insulator layer (MI) 
$n(x)=1$ and band insulator 
layer (BI) $n(x)=2$,
but varies in the 
metallic region (M). 
The width of MI layer is 
nonzero when $V_{\rm D}<-\delta/\kappa$ 
\begin{equation}
d_{\rm MI}=\sqrt{2\ve(-V_{\rm D}-\delta/\kappa)/e \delta}
\label{eq:MIwidth}
\end{equation} 
and saturates at $V_{\rm D}=-\delta/\kappa-\Delta$.
When the MI layer saturate,
the metallic region appears whose width is
\begin{widetext}
\begin{equation}
d_{\rm M}=\sqrt{\ve/e\kappa}
\cosh^{-1}\left[\left\{
(V_{\rm D}+\Delta)\delta/\kappa+
\sqrt{2\Delta\delta/\kappa}\sqrt{
(V_{\rm D}+\Delta+\delta/\kappa)^2
-2(\delta/\kappa)(V_{\rm D}+\delta/\kappa
)}\right\}/\{
2\Delta\delta/\kappa-(\delta/\kappa)^2\}
\right],
\label{eq:Mwidth}
\end{equation}
\end{widetext}
which saturates when $V_{\rm D}<-\frac{1+\delta}{\kappa}-\Delta$.
Finally, the width of the 
BI layer is nonzero when $V_{\rm D}<-\frac{1+\delta}{\kappa}-\Delta$
\begin{eqnarray}
d_{\rm BI}=\frac{-F_2+\sqrt{(F_2)^2-2\frac{e}{\ve}(1+\delta)
\left((1+\delta)/\kappa+V_{\rm D}+\Delta\right)
}}
{\frac{e}{\ve}(1+\delta)}
\label{eq:BIwidth}
\end{eqnarray}
with $F_1= \sqrt{2e\Delta\delta/\ve}$
and $F_2=F_1
\cosh\left(\sqrt{e\kappa/\ve}d_{\rm M}\right)+
\sqrt{e\kappa/\ve}
\left(\frac{\delta}{\kappa}\right)
\sinh\left(\sqrt{e\kappa/\ve}d_{\rm M}\right)$
where $d_{\rm M}$ here is obtained by 
substituting 
$V_{\rm D}=-\frac{1+\delta}{\kappa}-\Delta$ in eq. (\ref{eq:Mwidth}).
In Fig. \ref{fig:NvsV} (b), we plot the $V_{\rm D}$ dependence 
of the widths and compare them with the DMRG results,
and both agrees surprisingly well. 
This agreement is highly nontrivial; In the insulating phase
the localization length $\xi\sim W/\Delta(U)$ is of the order of
few sites, but in the metallic phase it should diverge. 
So, in the metallic phase the local approximation is not 
a priori justified. However, our numerical calculation
shows that it works remarkably well.
Based on this success, we apply this local 
approximation to more generic cases below.

\begin{figure}[thb]
\centering 
\includegraphics[width=7.cm]{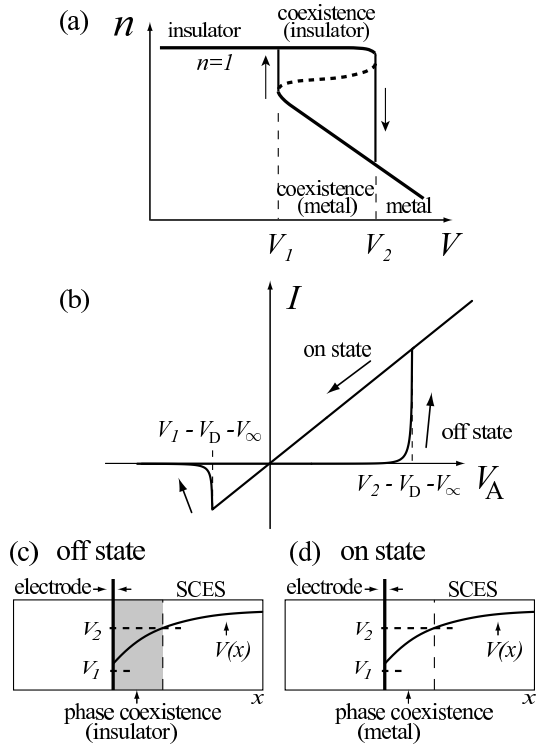}
\caption{
(a) Density-potential relation near Mott's transition
in dimensions higher than one (schematic). 
A region ($V_1<V<V_2$) exists where 
metallic and insulating phases coexist.
(b) $I$-$V$ characteristics 
of the metal/SCES interface when 
$V_1<V_{\rm D}+V_{\infty}<V_2$ is satisfied.
(c), (d) Pictures of the off state (c) and on state (d).
}
\label{fig:cer}
\end{figure}

{\it Application to CER ---}
In CER devices, the interface between the electrode and SCES is two dimensional. 
In systems with dimensions higher than 
one, the filling driven Mott transition 
is first order and hysteretic behavior
takes place near the transition point \cite{Fisher1995,Kotliar2002,SOnodaGL2004}.
We expect similar hysteresis in the density-potential 
relation of the interface Mott transition (Fig. \ref{fig:cer} (a)).
If this is the case, coexistence of the metallic state
and the insulating state is possible. 
Then, the $I$-$V$ characteristics of the 
device shows high non-linearity as well as a 
memory effect, i.e., RRAM behavior, when 
a layer of coexistence state exist in the SCES region.
The condition for this to take place
is $V_{\rm D}<-(\delta-\Delta\delta_1)/\kappa$.
For example, if $V_1<V_{\rm D}+V_{\infty}<V_2$ is satisfied,
the region nearest to the interface is 
a phase coexistence layer, as shown in Fig. \ref{fig:cer}
(c) and (d). If this is the case, the junction 
may be insulating (c) or metallic (d). 
They correspond to the off and
on states, respectively.
We can switch between them by 
applying a voltage $V_{\rm A}$ on the 
electrode forming a
hysteresis loop in the $I$-$V$ characteristics  Fig. \ref{fig:cer} (b). 
We assume a tunneling form for the current 
\begin{equation}
I(V_{\rm A})=T(V_{\rm A})I_0(V_{\rm A}),
\end{equation}
where $T(V_{\rm A})$
is the tunneling probability of the barrier
and $I_0(V_{\rm A})$ the current 
at the metal/SCES junction without any barrier.
$I_0(V_{\rm A})$ reflects the 
details of the device
and may be Ohmic, i.e.
$I_0(V_{\rm A})\propto V_{\rm A}$,
or if space charge limited current is realized, 
$I_0(V_{\rm A})\propto (V_{\rm A})^2$
\cite{LampertBook}. 
We assume that the tunneling probability decreases 
exponentially as the width of the MI layer grows,
i.e.,
$T(V_{\rm A})= e^{-d_{\rm MI}(V_{\rm A})/\xi}$, 
where $\xi$ is the decay length (here the temperature
dependence is neglected). 
Neglecting the jump of $n$ in Fig. \ref{fig:cer}
(a), the width of the MI layer
can be obtained 
by replacing $-V_{\rm D}-\delta/\kappa$ in eq.(\ref{eq:MIwidth}) 
with $V_{i}-V_{\rm D}-V_{\infty}-V_{\rm A},\;i=1,2$
for the on and off state, respectively. 
Thus, a hysteresis loop is realized
in the $I$-$V$ characteristics
as in Fig. \ref{fig:cer} (b). 
If we define the CER ratio by $\Delta R/R\equiv (R_{\rm off}-R_{\rm on})/R_{\rm on}$,
where $R_{\rm on,off}$ is the resistivity of the on and off states, 
we obtain $\Delta R/R = e^{\sqrt{2\ve(V_2-V_{\rm D}-V_\infty)/e\delta}/\xi}-1$. 
In a Ti/PCMO based CER device, however,
neither the on nor the off states 
show Ohmic $I$-$V$ characteristics \cite{Sawa2004}.
This can be also understood by our model with 
$V_{\rm D}+V_\infty<V_1$. 
In such cases, the CER ratio becomes 
$\Delta R/R=e^{\left(\sqrt{2\ve(V_2-V_{\rm D}-V_\infty)/e\delta}
-\sqrt{2\ve(V_1-V_{\rm D}-V_\infty)/e\delta}\right)/\xi}-1$.
In either case, we can design a CER device with larger CER ratio by 
decreasing the doping ratio $\delta$ and making the 
phase coexistence region wider.

In summary, we have studied the interface Mott transition 
by the DMRG method and by Poisson's equation combined 
with a local equilibrium ansatz. 
We proposed a novel mechanism of CER for materials with a 
first order metal-insulator transition.

We thank A. Sawa, I. H. Inoue, M. Kawasaki
and Y. Tokura for illuminating discussions,
and Y. Ogimoto for careful reading of the manuscript. 
TO acknowledges S. Onoda, and F. Ogushi for helpful comments.


\end{document}